# Efficient Implementation of Elliptic Curve Cryptography Using Low-power Digital Signal Processor

Muhammad Yasir Malik
*National University of Science and Technology (NUST), Pakistan*
yasir_alf@yahoo.com

*Abstract*— RSA (Rivest, Shamir and Adleman) is being used as a public key exchange and key agreement tool for many years. Due to large numbers involved in RSA, there is need for more efficient methods in implementation for public key cryptosystems. Elliptic Curve Cryptography (ECC) is based on elliptic curves defined over a finite field. Elliptic curve cryptosystems (ECC) were discovered by Victor Miller [1] and Neal Koblitz [2] in 1985. This paper comprises of five sections. Section I is introduction to ECC and its components. Section II describes advantages of ECC schemes and its comparison with RSA. Section III is about some of the applications of ECC. Section IV gives some embedded implementations of ECC. Section V contains ECC implementation on fixed point Digital Signal Processor (TMS320VC5416). ECC was implemented using general purpose microcontrollers and Field Programmable Gate Arrays (FPGA) before this work. DSP is more powerful than microcontrollers and much economical than FPGA. So this implementation can be efficiently utilized in low-power applications.

*Keywords*— Efficient Implementation of ECC, Elliptic Curve Discrete Logarithmic Problem, Advantages of ECC, Scalar multiplication, Montgomery modular multiplication, Low-power, Digital signal processor

## I. INTRODUCTION

Strength of RSA [3] lies in integer factorization problem. That is when we are given a number n; we have to find its prime factors. It becomes quite complicated when dealing with large numbers. This is the strength of RSA and to an extent, is the disadvantage associated with it.

Elliptic curve is a curve that is a group. ECC utilizes this group for its functioning. Its strength is the problem involving elliptic curves; Elliptic Curve Discrete Logarithmic Problem (ECDLP). That is when an elliptic curve E and points P and Q on E are given, find 'x' when Q=xP.

A simple elliptic curve with points is shown in Figure 1.

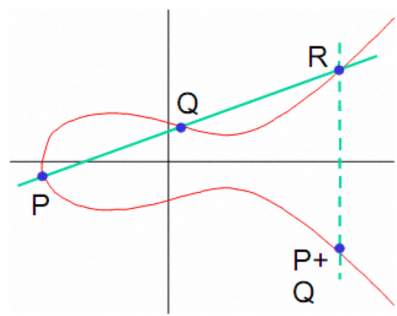

Figure 1. A simple elliptic curve

As described earlier the disadvantage of RSA is the use of large numbers (and hence large keys) for its operation. More secure we want a RSA public key cryptosystem to be; the larger would be the numbers involved. This will increase the key size to a large extent. If we don't want to compromise on the security of our information then we have to use a more efficient method involving less number of bits (and lesser key size). This new cryptosystem must be strong enough to ensure same or even greater level of security for our information.

Based on these concerns ECC clearly stands out as a much better and efficient method for public key cryptosystems. The novel idea behind its application, the strength of ECDLP and its much easier and memory efficient implementation makes it the choice of the present and new public key protocols and systems. SO ECC can be called the future generation of public key systems.

Fundamentals of ECC are given comprising of basic equations, fields used in ECC and the problem on which these systems rely.

### A. Basics of ECC

An elliptic curve 'E' is a curve given by an equation (for a cubic or quadratic polynomial f(x)):

$$E: y2 = f(x) \qquad (1)$$

We want that the polynomial f(x) has no double roots to ensure that the curve is non-singular.

After a change of variables, the equation takes the simpler form (cubic):

$$E: y2 = x3 + a\,x + b \quad (2)$$

Extra point Θ "at infinity" is added to the above equation so that E is really the set.

$$E = \{(x,y): y2 = x3 + a\,x + b\} \cup \{\Theta\} \quad (3)$$

If we have a point $P_1(x_1,y_1)$ on any elliptic curve and we want to find $P_2(x_2,y_2)$ such that $P_2 = 2P_1$. This is known as Point Doubling and it can be done as:

$$Let\ \lambda = x1 + \frac{y1}{x1}, then\ x2 = a + \lambda + \lambda^2$$

$$and\ y2 = (x1 + x2)\lambda + x2 + y1 \quad (4)$$

If we have two points $P_1(x_1,y_1)$ and $P_2(x_2,y_2)$ on any elliptic curve and we want to find $P_3(x_3,y_3)$ such that $P_3 = P_1 + P_2$. This is known as Point Addition and it can be done as:

$$Let\ \lambda = \frac{y1+y2}{x1+x2}, then\ x3 = a + \lambda + \lambda^2 + x1 + x2$$

$$and\ y3 = (x2 + x3)\lambda + x3 + x2 \quad (5)$$

ECC involves elliptic curves defined over a finite field. There are two types of fields of interest
- Prime fields GF(p)
- Binary finite fields GF(2^m)

Points on the elliptic curve is written as P(x,y) where x and y are elements of GF(p).

The size of a set of elliptic curve domain parameters on a prime curve is defined as the number of bits in the binary representation of the field order; commonly denoted p. Size on a characteristic-2 curve is defined as the number of bits in the binary representation of the field, commonly denoted as m.

### B. Elliptic Curve Discrete Logarithmic Problem (ECDLP)

ECDLP has following components:
- A well defined finite field GF(p) or GF(2^m). An elliptic curve E defined over any of these two defined finite fields
- A point P, of higher order, present on elliptic curve E
- A scalar multiple of P, let's say k, such that k.P=P+P+P+…+P (k times)

So ECDLP involves scalar multiplication. Now when we have k and P then it is quite easier to find k.P. But when we have to find k for given P and k.P, the task is bit studious.

## II. ADVANTAGES OF ECC

Some of the advantages that come with ECC systems are briefly explained here.

### A. More Complex

In spite of multiplication or exponentiation in finite field, ECC uses scalar multiplication. Solving Q=k.P (utilized by ECC) is more difficult than solving factorization (used by RSA) and discrete logarithm (used by Diffie-Hellman (DH), ElGamal, Digital Signature Algorithm (DSA)). So ECC is much stronger than other public key agreement and signature authentication methods.

### B. Involvement of Less Number of Bits

ECC requires much lesser numbers (and thus less number of bits) for its operation thanks to ECDLP. The security level of a 160-bit ECC, 1024-bit RSA, and (160/1024)-bit DSA are similar. Table 1 shows the comparison between ECC and RSA.

Table 1. ECC vs. RSA

| ECC (bits) | RSA (bits) | Key size ratio | AES (bits) |
|---|---|---|---|
| 160 | 1024 | 1:6 | -- |
| 256 | 3024 | 1:12 | 128 |
| 384 | 7680 | 1:20 | 192 |
| 512 | 16,360 | 1:30 | 256 |

### C. Wide Selection of Finite Fields and Curves

Different finite fields can be used for ECC according to security requirements. Finite fields which can be used for ECC are defined in Standards for Efficient Cryptography1 [4].

For GF (p) the finite fields used can be from the following defined set:
p ε {112; 128; 160; 192; 224; 256; 384; 512; 1024}

For GF (2^m) the finite fields used can be from the following defined set:
m ε {113; 131; 163; 193; 233; 239; 283; 409; 571}

Many different curves can be chosen for the same field by different users. Many such curves and their domain parameters are defined in Standards for Efficient Cryptography2 [5].

### D. Power Consumption

ECC requires less power for its functioning so it is more suitable for low power applications such as handheld and mobile devices.

### E. Computational Efficiency

Implementing scalar multiplication in software and hardware is much more feasible than performing multiplications or exponentiations in them. As ECC makes use of scalar multiplications so it is much more computationally efficient than RSA and Diffie-Hellman (DH) public schemes.

So we can say without any doubt that ECC is the stronger and the faster (efficient) amongst the present techniques.

## III. APPLICATIONS OF ECC

Due to its small key sizes ECC is becoming a widely utilized and attractive public-key cryptosystem. Compared to cryptosystems such as RSA, DSA, and DH, ECC variations on these schemes offer equivalent security with smaller key sizes. This is illustrated in Table 2. L is size of field, N is sub-field size.

Table 2. Comparable Key Sizes (In Bits)

| Symmetric | Discrete Log (DSA, DH) | RSA | ECC |
|---|---|---|---|
| 80 | L = 1024 N = 160 | 1024 | 160-233 |
| 112 | L = 2048 N = 256 | 2048 | 224-255 |
| 128 | L = 3072 N = 256 | 3078 | 256-383 |

| Symmetric | Discrete Log (DSA, DH) | RSA | ECC |
|---|---|---|---|
| 192 | L = 7680 N = 384 | 7680 | 384-511 |
| 256 | L = 15360 N = 512 | 15360 | 512+ |

Smaller key sizes result in less power, bandwidth, and computational requirements. This makes ECC a good choice for low power environments. ECC has got applications as a public key sharing scheme and as digital signature authentication scheme. The applications of ECC are:
   A. Elliptic Curve Diffie-Hellman (ECDH) Key Exchange
   B. Elliptic Curve Menezes-Qu-Vanstone (ECMQV) Key Exchange and Verification
   C. Elliptic Curve Digital Signature Algorithm (ECDSA)

*A. Elliptic Curve Diffie-Hellman (ECDH) Key Exchange*

ECDH operates by providing the two parties sharing a secret key with a public key, which in this case is a point P on elliptic curve E. Alice performs scalar multiplication using this point P and a scalar multiple a, which is secret key of Alice. a.P now becomes public key of Alice which she can share with the other party.

On the other end, Bob performs scalar multiplication using point P and a scalar multiple of his choice i.e. b, which is secret key of Bob. b.P becomes public key of Bob which he shares with Alice.

Alice performs scalar multiplication of public key of Alice (b.P) with her secret key a to get a.b.P. Bob also does the same with his secret key b and public key of Alice a.P to get the same a.b.P. This entity i.e. a.b.P is same for both the parties and is their shared key.

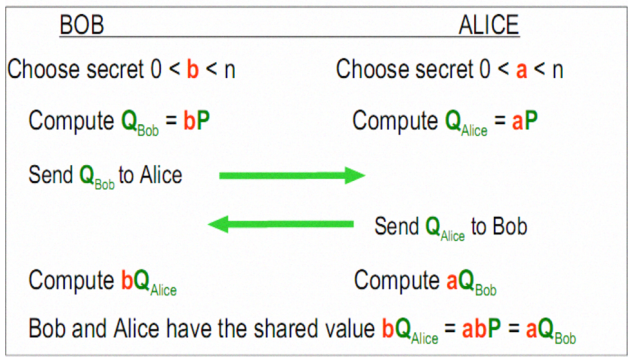

Figure 2. Elliptic Curve Diffie-Hellman (ECDH)

*B. Elliptic Curve Menezes-Qu-Vanstone (ECMQV) Key Exchange and Verification*

ECMQV key exchange algorithm is used to generate a shared secret key from two elliptic curve key pairs owned by one entity and two elliptic curve public keys owned by another entity. Both entities have the analogous role in the algorithm. Each entity makes use of its key pairs and other entity's public keys to get its secret key. As the secret key obtained by each side by using this algorithm is the same, key exchange and verification is achieved by this process.

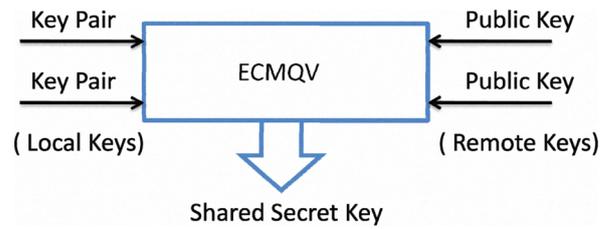

Figure 3. ECMQV

*C. Elliptic Curve Digital Signature Algorithm (ECDSA)*

A digital signature is a number dependent on some secret known only to the signer (the signer's private key), and additionally on the contents of the message being signed.

ECDSA is the elliptic curve analogue of the Digital Signature Algorithm (DSA). Key pair in ECDSA is generated the same way as of that in ECDH. ECDSA uses hashing of message and operations on points to generate signatures.

## IV. HARDWARE IMPLEMENTATIONS

In this section some of the notable implementations of ECC on general purpose microcontrollers and FPGA are given. Public key cryptography involves large numbers and hence is considered to be slow. Most of the public-key cryptography is implemented on small devices in conjunction with special purpose cryptographic hardware. Accelerators for many crypto functions are used along with small processors.

However in [6], authors implemented ECC without any special hardware. With the help of their new algorithm that reduces memory accesses, they achieved 160-bit ECC point multiplication on an Atmel ATmega128 at 8 MHz at 0.81s. That is the best known time for such an operation without using specialized hardware.

Software and hardware co-design of ECC {GF ($2^{191}$)} was implemented in [7] using Dalton 8051 and special hardware. The hardware part consists of an elliptic curve acceleration unit (ECAU) and an interface with direct memory access (DMA) to enable fast data transfer between the ECAU and the external RAM (XRAM) attached to the 8051 microcontroller.

The ECAU allows to perform a full scalar multiplication over the field GF ($2^{191}$) in about 118 msec, assuming that the Dalton 8051 is clocked with 12 MHz, a scalar multiplication over the field GF ($2^{163}$) takes less than 100 msec. System block diagram for this configuration is shown in Figure 4.

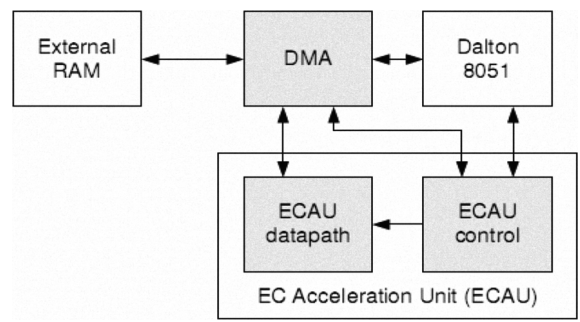

Figure 4. System block diagram

Many implementations of ECC on FPGA are given. The time taken by them varies because of different types of implementations with variants in techniques applied. Timings for finite field/ECC operations of some hardware platforms are given in Table 3.

Table 3. Timings of ECC Operations

| FPGA/ASIC | Operation | Time | References |
|---|---|---|---|
| FPGA | 512bit multiplication | 2.37ms | [8] |
| FPGA | 163bit ECC scalar mult. | 144s | [9] |
| FPGA | 191bit ECC scalar mult. | 3ms | [10] |
| ASIC | 160 bit multiplication GF(p) | 4.1s | [11] |

## V. IMPLEMENTATION USING 54xx DSP

This section contains implementation of ECC by using platform of Digital Signal Processors (DSP). TMS320VC5416 Fixed-Point Digital Signal Processor (Texas Instruments) is used for this purpose. It is a 16-bit processor with operations at 160 MHz. It incorporates 128K x 16-Bit On-Chip RAM and 16K × 16-Bit On-Chip ROM which is configured for Program Memory [12].

As mentioned earlier there are two variants of the finite fields which are defined for ECC. One is prime finite field and the other is binary field. ECC makes use of both of these fields. It depends on our needs and choice. According to the requirements of security, power and other factors, we can chose from available sets of these fields. I have used prime field and chose 160-bits for my implementations. According to [13], even if we follow Moore's law ECC-160 (or ECC-160p) will remain completely secure till 2020!

I used the efficient and reduced method mentioned in [14] for my execution of ECC-160 bit.

### A. Components of ECC Implementation

The components required for executing the task are defined here:
- A prime number p
- A point P (with its components x and y) on a defined elliptic curve
- A scalar multiple k
- Let 'b' be character base and it depends on the number of bits of processor. As 54xx is a 16-bit processor so b in this case is $2^{16}$
- We select a positive integer R which is larger than p and co-prime with p
  We may use $R=b^t > p$. t is taken as 10. So R in this case is $2^{160}$

Values of prime number p, point P and scalar multiple k used in my implementation are from the Recommended Domain Parameters document [5].

### B. Conversion to Jacobian Coordinates

Test vectors defined in [15] are used. We take a point P on the chosen elliptic curve and represent P as P(x,y). First of all we define the components of P (which are x and y) in Jacobian coordinate system. Here z is taken as 1.

### C. Conversion of General Numbers into Montgomery Numbers

As ECC requires multiplying with large numbers so we use Montgomery multiplication for fulfill the purpose efficiently. This helps in ease of processes and computation s thus reducing time and resources required in completion of ECC.

Data points (x, y, z) are converted into Montgomery numbers.

Let $n = 16 * (10)$ for ECC-160
Input x
Output $X = x * R$ (modulo p)
1. $X = x$
2. i from 0 to n-1
   2.1 $X = X * 2$
   2.2 If $X \geq p$, then $X = X - p$
   2.3 $i = i + 1$. Go to step 2
3. Output X                                                      (6)

### D. Computation of Montgomery Modular Multiplication

All the multiplications performed in the process are the Montgomery multiplications. They follow the following algorithm.

$pbar = -p^{-1}$ (modulo b)
Inputs: X and Y (They are Montgomery converted x and y)
Output: $Z = XYR^{-1}$ (modulo p)
1. Let $Z = 0$
2. Repeat the steps of j from 0 to t-1
   2.1 $u = (z_0 + x_j y_0)$ pbar (modulo b)
   2.2 $Z = (Z + x_j y + up)/b$
   2.3 $j = j + 1$
3. If $Z \geq p$, then $Z = Z - p$
4. Output Z                                                      (7)

### E. Scalar Multiplication

In scalar multiplication we have a scalar multiple which we multiply with the point P. In ECC-160, this scalar multiple is a 160 bit number. This number acts as secret keys of different entities.

In ECC we perform point addition and point doubling based upon the value of the bits of k. The algorithm for scalar multiplication (or point multiplication) can be written as:

Input: Point P components (X,Y,Z) and scalar multiple k
Output: k.P
1. Check value of k from its MSB to position 0 (MSB is the most significant bit of k)
2. If checked bit = 0,
   - Perform Point Double on data points
3. If checked bit = 1,
   - Perform Point Double on data points
   - Perform Point Addition on data points

*4. Output k.P* (8)

### F. Conversion to Affine Coordinates

General numbers are then converted from Jacobian coordinates to Affine Coordinates. This is achieved by:

$$x = x/z^2, y = y/z^3 \qquad (9)$$

### G. Restoration to General Numbers

Now we have performed point multiplication, we need to restore the number we get (k.P) back to general number. Montgomery number can be restored back by using the following algorithm:

*Input: X (Montgomery number)*
*Output: $x = XR^{-1}$ (modulo p)*
*1. $x = X$*
*2. Repeat the steps from i = 0 to n-1*
  *2.1 $x = x/2$ if x is an even number*
  *2.2 Otherwise $x = (x + p)/2$*
  *2.3 i = i + 1; Go to step 2*
*3. Output x (General number)* (10)

After restoration we have achieved our goal of point multiplication in ECC-160.

### H. Timings of Different Operations in ECC

Time taken by different processes (approx.) during ECC-160 bit on 54xx DSP is given Table 4.

**Table 4. Timings of Operations in ECC**

| Operation | CPU cycles | Time taken |
|---|---|---|
| Addition in GF(p) | 315 | 1.97 us |
| Subtraction in GF(p) | 357 | 2.23 us |
| Montgomery Multiplication | 2,860 | 17.88 us |
| Point Addition | 33,049 | 207 us |
| Point Doubling | 40,737 | 254 us |
| Scalar Multiplication | 10,148,863 | 63.4 ms |

## VI. CONCLUSIONS

This implementation takes only fraction of a second. Along with time consumption ECC provides power consumption too. This makes it an ideal choice for portable, mobile and low power applications. It can be a very secure and useful replacement of already being used cryptosystems for key exchange, key agreement and mutual authentication (like RSA, ElGamal and their variants).

Its use can be further extended in smart cards and RFID applications because of less memory and low power requirements.

### REFERENCES

[1] V. S. Miller, *Use of Elliptic Curves in Cryptography, Advances in Cryptology*, Vol. 218, pp. 417-426, 1985.
[2] N. Koblitz, Elliptic Curve Cryptosystems, *Mathematics of Computation*, Vol. 48, pp. 203-209, 1987.
[3] R. L. Rivest, A. Shamir, and L. Adleman. *A Method for Obtaining Digital Signatures and Public-Key Cryptosystems*. Communications of the ACM, 21(2):120–126, February 1978.
[4] STANDARDS FOR EFFICIENT CRYPTOGRAPHY, SEC 1: Elliptic Curve Cryptography, Certicom Research, September 20, 2000. Version 1.0
[5] STANDARDS FOR EFFICIENT CRYPTOGRAPHY, SEC 2: Recommended Elliptic Curve Domain Parameters, September 20, 2000. Version 1.0
[6] Nils Gura, Arun Patel, Arvinderpal Wander, Hans Eberle, Sheueling Chang Shantz, *Comparing Elliptic Curve Cryptography and RSA on 8-bit CPUs*. Can be found at: http://www.research.sun.com/projects/crypto
[7] Manuel Koschuch, Joachim Lechner, Andreas Weitzer, Johann Großsch¨adl, Alexander Szekely, Stefan Tillich, and Johannes Wolkerstorfer, *Hardware/Software Co-Design of Elliptic Curve Cryptography on an 8051 Microcontroller*, CHES 2006, LNCS 4249, pp. 430–444, 2006.
[8] T. Blum and C. Paar, *Montgomery modular multiplication on reconfigurable hardware*. In Proceedings of the 14th IEEE Symposium on Computer Arithmetic (ARITH-14), pages 70–77, 1999.
[9] N. Gura, S. Chang, H. Eberle, G. Sumit, V. Gupta, D. Finchelstein, E. Goupy, and D. Stebila. *An End-to-End Systems Approach to Elliptic Curve Cryptography*. In Ç. K. Koç and C. Paar, editors, Cryptographic Hardware and Embedded Systems — CHES 2001, volume LNCS 1965, pages 351–366. Springer-Verlag, 2001.
[10] G. Orlando and C. Paar, *A Scalable GF(p) Elliptic Curve Processor Architecture for Programmable Hardware*. In Ç. K. Koç, D. Naccache, and C. Paar, editors, Workshop on Cryptographic Hardware and Embedded Systems — CHES 2001, volume LNCS 2162, pages 348–363. Springer-Verlag, May 14-16, 2001.
[11] E. Savas, A. F. Tenca, and C . K. Koç. *A scalable and unified multiplier architecture for finite fields gf(p ) and gf(2)*. In Ç. K. Koç and C. Paar, editors, Cryptographic Hardware and Embedded Systems — CHES 2000, volume LNCS 1965, pages 281–296. Springer-Verlag, 2000.
[12] *TMS320VC5416 Fixed-Point Digital Signal Processor, Data Manual*, Literature Number: SPRS095J, March 1999 – Revised April 2003.
[13] Sandeep Kumar , Christof Paar , Jan Pelzl , Gerd Pfeiffer , and Manfred Schimmler, *Breaking Ciphers with COPACOBANA- A Cost-Optimized Parallel Code Breaker*, CHES 2006, Yokohama, October 2006.
[14] Darrel Hankerson, Alfred Menezes, and Scott Vanstone, *Guide to Elliptic Curve Cryptography*, Springer, 2004.
[15] GUIDELINES FOR EFFICIENT CRYPTOGRAPHY. GEC2: TestVectors for SEC1. Working Draft, September, 1999. Version 0.3
[16] M. Morales-Sandoval and C. Feregrino-Uribe, *On the hardware design of an elliptic curve cryptosystem*, Proceedings of the Fifth Mexican International Conference in Computer Science, pp. 60-70, 2004.
[17] Guerric Meurice de Dormale and Jean-Jacques Quisquater, *High-speed hardware implementations of Elliptic Curve Cryptography: A survey*, J. Syst. Archit., Vol. 53, pp. 72-84, 2007.
[18] H. Aigner, H. Bock, M. H¨utter, and J. Wolkerstorfer. *A low-cost ECC coprocessor for smartcard*, In Cryptographic Hardware and Embedded Systems — CHES 2004, LNCS 3156, pp. 107–118. Springer Verlag, 2004.
[19] Jan Pelzl, *Hardware Implementation of ECC*, Can be found at: http://www.crypto.rub.de